\documentclass[prl,preprintnumbers,superscriptaddress,twocolumn]{revtex4-2}
\usepackage{graphicx}
\usepackage{bm}
\usepackage{amsmath,amssymb,amsfonts,textcomp}
\usepackage{multirow}
\usepackage[colorlinks=true,linkcolor=blue,citecolor=blue,urlcolor=blue]{hyperref}
\usepackage{tabularx}
\usepackage{gensymb}
\usepackage{color}
\usepackage{array}
\usepackage{hhline}
\usepackage[dvipsnames]{xcolor}
\usepackage[normalem]{ulem}
\usepackage{verbatim}   
\usepackage{subfigure}  
\usepackage{braket}
\usepackage{float}
\usepackage{subfigure}
\usepackage{graphicx}
\begin{document}

\title{Bond-dependent interactions and ill-ordered state \\
in the honeycomb cobaltate BaCo$_{2}$(AsO$_{4}$)$_{2}$}

\author{A. Devillez}
\author{J. Robert}
\author{E. Lhotel}
\author{R. Ballou}
\author{C. Cavenel}
\author{F. Denis Romero}
\affiliation{Universit\'e Grenoble Alpes, CNRS, Institut N\'eel, 38042 Grenoble, France}


\author{Q. Faure}
\affiliation{Laboratoire L\'eon Brillouin, CEA, CNRS, Universit\'e Paris-Saclay, CEA-Saclay, 91191 Gif-sur-Yvette, France}


\author{H. Jacobsen}
\affiliation{PSI Center for Neutron and Muon Sciences, 5232 Villigen PSI, Switzerland}
\affiliation{Nanoscience Center, Niels Bohr Institute, University of Copenhagen, 2100 Copenhagen, Denmark}

\author{J. Lass}
\author{D. G. Mazzone}
\affiliation{PSI Center for Neutron and Muon Sciences, 5232 Villigen PSI, Switzerland}

\author{U. Bengaard Hansen}
\author{M. Enderle}
\affiliation{Institut Laue-Langevin, 71 avenue des Martyrs CS 20156, 38042 Grenoble Cedex 9, France}

\author{S. Raymond}
\affiliation{Universit\'e Grenoble Alpes, CEA, IRIG, MEM, MDN, 38042 Grenoble, France}

\author{S. De Brion}
\author{V. Simonet}
\author{M. Songvilay}
\affiliation{Universit\'e Grenoble Alpes, CNRS, Institut N\'eel, 38042 Grenoble, France}

\date{\today}

\begin{abstract}
The ground state and Hamiltonian of the honeycomb lattice material BaCo$_{2}$(AsO$_{4}$)$_{2}$ hosting magnetic Co$^{2+}$, have been debated for decades. The recent proposal for anisotropic bond-dependent interactions in such honeycomb cobaltates has raised the prospect of revisiting its Hamiltonian in the context of Kitaev physics. To test this hypothesis, we have combined magnetization, ac-susceptibility and neutron scattering measurements on a BaCo$_{2}$(AsO$_{4}$)$_{2}$ single-crystal, together with advanced modeling. Our experimental results highlight a collinear magnetic ground state with intrinsic disorder associated to an average incommensurate propagation vector. Monte Carlo simulations and linear spin wave calculations were performed to obtain a spin model compatible with this unusual ground state, the dispersion of magnetic excitations and a magnetization plateau under magnetic field. We thus show that bond-dependent anisotropic interactions, including Kitaev-like interactions, are necessary to account for the puzzling properties of this long-explored material,  and are hence a general ingredient in the cobaltates.

\end{abstract}


\maketitle

\begin{figure}
\includegraphics[scale=0.33]{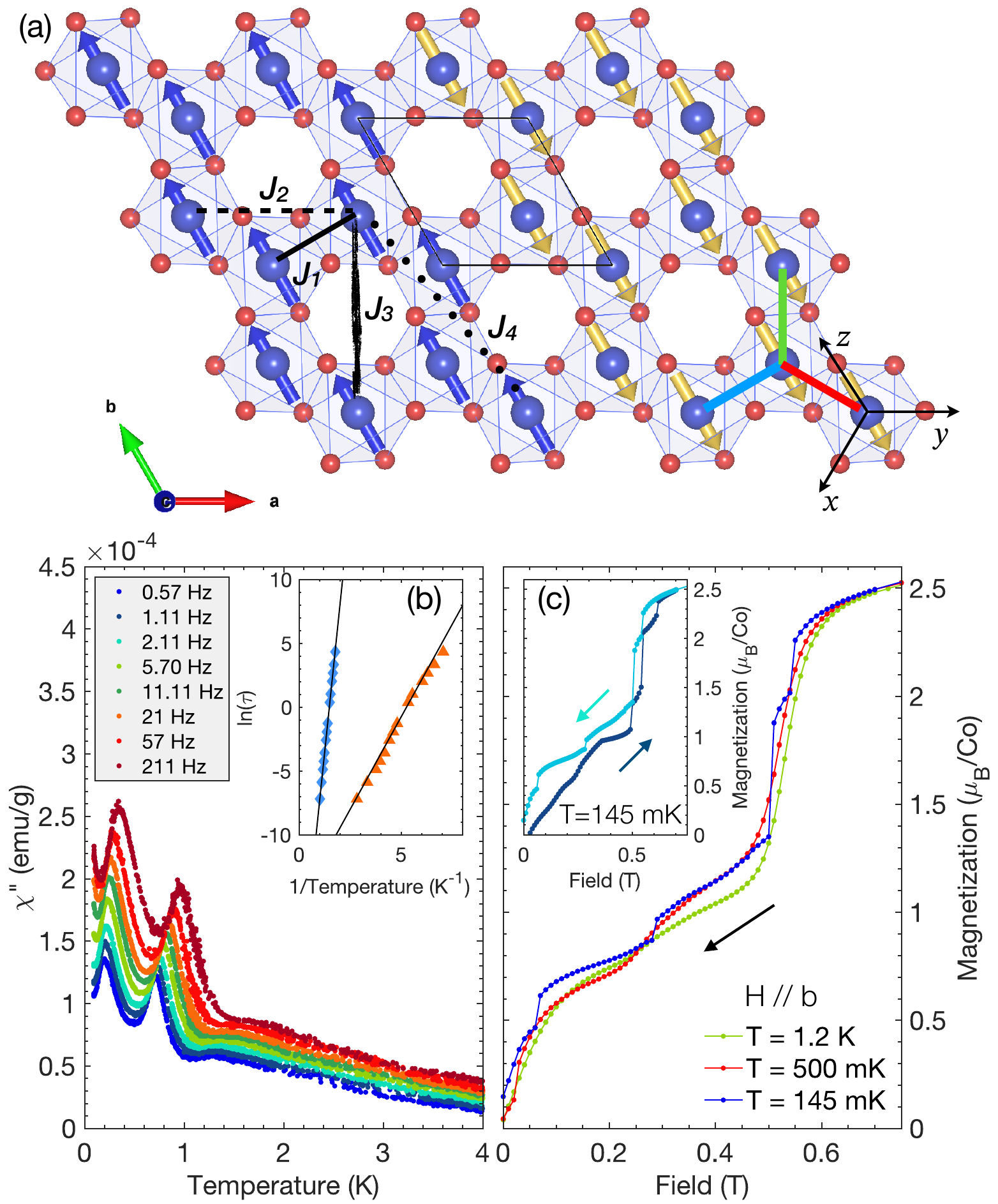}
\caption{ \label{fig:chi_ac} (a) Honeycomb plane in BCAO, with the proposed $uudd$ ordered sequence as the basing block of the magnetic structure. The considered Heisenberg exchange interactions are displayed in black. The red, green and blue lines refer to the $x$, $y$, and $z$ bonds respectively, for the anisotropic interactions defined in the local $xyz$ orthogonal frame. (b) Imaginary part of the ac susceptibility $\chi''$ plotted as a function of temperature for various frequencies  with $H_{\rm ac}=3.3$ Oe. The inset displays the relaxation times $\tau$ associated to  the low and high temperature $\chi''$ peaks (in blue and orange respectively) as a function of inverse temperature. The lines represent fits to Arrhenius laws. (c) Magnetization curves as a function of decreasing magnetic field, measured at T = 1.2 K, 500 mK and 145 mK. Inset : Hysteresis observed at 145 mK while for increasing and decreasing field. Measurements in (b) and (c) were performed with the field applied along ${\bf b}$.}
\end{figure}

In transition metal oxides, the interplay between crystal field, spin-orbit coupling, along with lattice geometry can induce a rich variety of magnetic properties. Recent focus has been drawn towards heavy transition metals with $4d$ and $5d$ electrons displaying a large spin-orbit coupling in particular thanks to the exciting proposal of Kitaev anisotropic bond-dependent interactions \cite{Kitaev2006,Jackeli2009}. This has led to numerous experimental and theoretical studies of a large panel of iridate/ruthenate materials hosting unconventional properties from potential spin liquids to topological and superconducting phases \cite{Rau2016, Schaffer2016}. Another interesting case is when magnetic interactions and spin-orbit coupling directly compete, as illustrated in $3d$ transition metals, which display weaker spin-orbit coupling but stronger electron correlations. Particularly rich examples are found in cobalt oxides, some of which have been studied for many years \cite{Regnault1977, Regnault1979, Regnault1983, Regnault2006, Regnault2018}. 
In the systems of interest, the Co$^{2+}$ ions host a sufficiently large spin-orbit coupling to stabilize a spin-orbital $j_{eff}$~=~1/2 Kramers ground state and  it has been argued that, in contrast to the iridate and ruthenate counterparts, due to the presence of additional spin-active $e_g$ electrons in Co$^{2+}$, the Heisenberg term can be suppressed to the profit of Kitaev interactions \cite{Liu2018, Sano2018, Liu2020, Liu2021,Yuan2020, Elliot2021, Coldea2010,Lefrancois2016,Wong2016,Bera2017,Songvilay2020,Chen2021,Yao2022,Kim2022,Sanders2022,Kruger2023}. 
This has led to the experimental evidence of Kitaev interactions in some cobaltate materials but always competing with other terms in the Hamiltonian such as Heisenberg, off-diagonal, further neighbor interactions, or single-ion anisotropy. 
Even though these additional ingredients push the investigated cobaltates away from a Kitaev spin liquid, the presence of anisotropic bond-dependent interactions produces interesting physics, in particular new magnetic ground states such as the original triple-q orders proposed in Na$_2$Co$_2$TeO$_6$ and Na$_3$Co$_2$SbO$_6$ \cite{Chen2021, Yao2022, Kruger2023,Garlea2024, Gu2024}.


Among the investigated cobaltates, BaCo$_{2}$(AsO$_{4}$)$_{2}$ (BCAO) is one example of a honeycomb magnet which has been extensively studied since the 80's \cite{Regnault1977, Regnault1979, Regnault1983, Regnault2006, Regnault2018,Halloran2023}, yet open questions remain about its magnetic ground state and excitations. Contradicting reports concerning its Hamiltonian, the presence or not of bond-dependent interactions, their influence on the properties, have been evoked recently. However, none of the proposed models seem to account for all observations, from the field-induced magnetization to the modeling of the spin wave dispersion in zero field. The latter, a stringent test to elucidate the physical parameters at play, has not been achieved yet.  

In this work, we address the long-standing issue around the nature of BCAO's magnetic ground state and excitations, and highlight the role of bond-dependent anisotropic interactions in stabilizing its original spin arrangement by combining neutron scattering and very low temperature magnetic measurements on a single-crystal, along with Monte-Carlo and linear spin wave calculations. Our results also questions the ubiquity of Kitaev-like interactions in cobaltate materials in general and their role in stabilizing new exotic spin textures.



BCAO crystallizes in the hexagonal R\={3} space group and comprises honeycomb layers of magnetic Co$^{2+}$ ions in the $(a,b)$ plane, with the inter-plane distance more than twice the intra-plane Co-Co distance, leading to a quasi-two-dimensional system. Previous neutron studies have shown that, in zero field, BCAO undergoes a sharp magnetic transition below T$_{N}$ $=$ 5.3~K, and orders with a propagation vector \textbf{k}~=~(0.27, 0, -4/3) \cite{Regnault1977, Regnault1983}. 
One particular unsettled issue concerns BCAO magnetic order: while first thought as a helical spin structure with XY anisotropy, neutron spherical polarimetry measurements  by L.-P. Regnault \textit{et al.} eventually suggested that the magnetic order may rather be collinear, made of ferromagnetic spin chains with magnetic moments aligned close to the $b$ direction, the chains being organized in an antiferromagnetic pattern along the $a^*$ direction (see Fig. \ref{fig:chi_ac}(a)). This pattern would correspond to an up-up-down-down $(uudd)$ configuration (also referred to as `double-zigzag") with inserted chain defects allowing to match the incommensurate propagation vector ($q_h\approx$~0.27 instead of 0.25 expected for a $uudd$ commensurate order). In favor of the presence of disorder, the chains were showed to be weakly coupled along the $a^*$ direction with anisotropic correlation lengths $\xi$ measured by neutron scattering within the honeycomb plane ($\xi_b/\xi_{a*}\approx2.7$) \cite{Regnault2006, Regnault2018}. Since this work, no further experimental evidence has confirmed this ill-ordered collinear magnetic state. 

Such a state is expected to display slow dynamics associated with the chain defects inserted in the $(uudd)$ arrangement, that can be probed by ac susceptibility measurements. Figure \ref{fig:chi_ac}(b) displays the imaginary part of the ac susceptibility $\chi^{''}$ measured as a function of temperature down to 100 mK and at various frequencies $f$ below T$_{N}$. Two peaks are observed, whose position varies with frequency, confirming that slow dynamics are present in the ordered state. As shown in the inset of Fig. \ref{fig:chi_ac}(b), the corresponding relaxation times $\tau=1/2\pi f$ follow Arrhenius laws $\tau = \tau_0 \exp{(E/T)}$, illustrating that both are associated with thermally activated processes over an energy barrier $E$, with a characteristic relaxation time $\tau_0$. The fits yield $E_{\rm LT}=2.8(1)$~K, $\tau_{\rm 0LT}=2.5(8)\times10^{-7}$~s and $E_{\rm HT}=15.5(1)$~K, $\tau_{\rm 0HT}\sim10^{-10}$~s for the low and high temperature peaks, respectively.

\begin{figure}
 \includegraphics[scale=0.36]{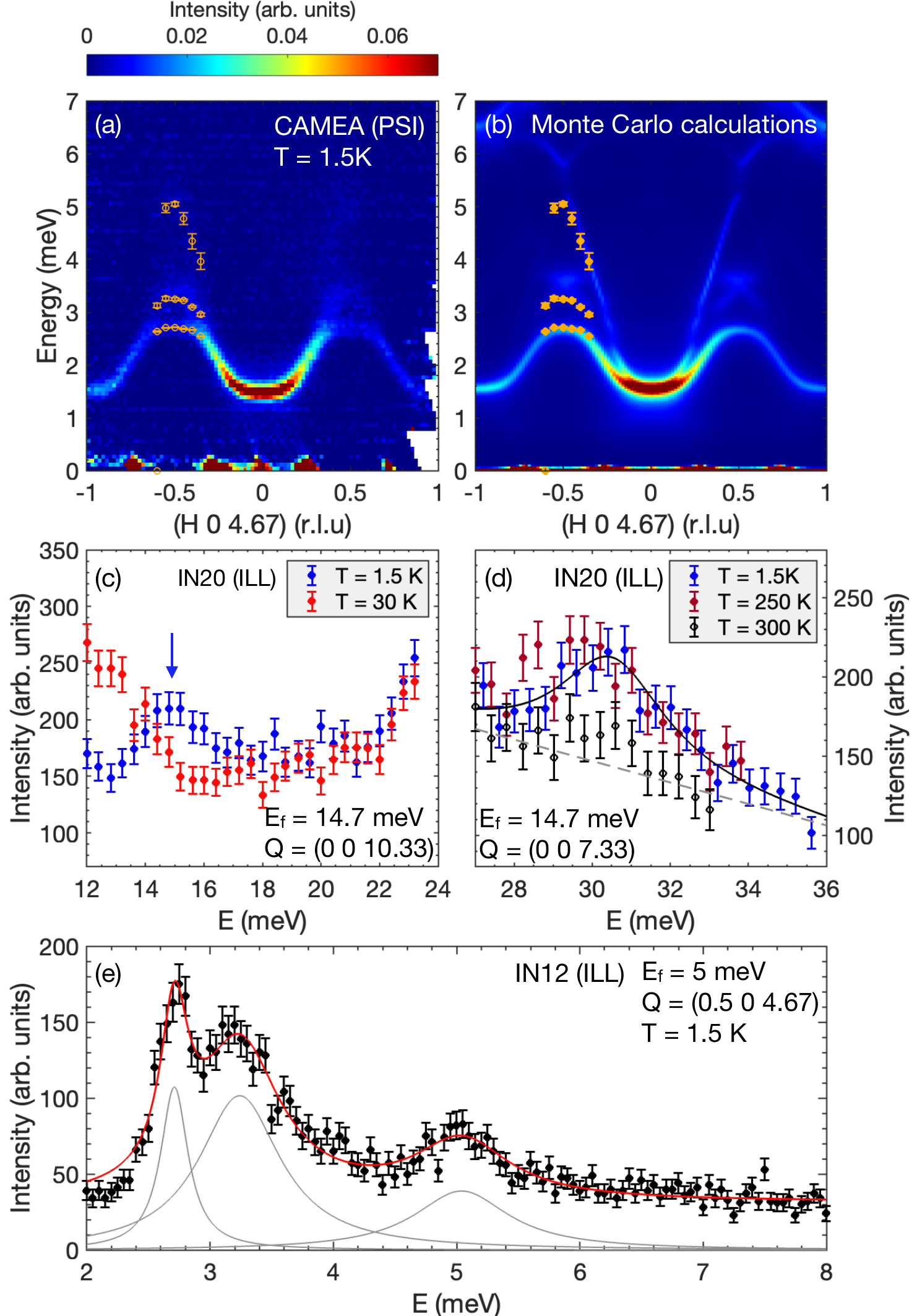}
 \caption{ \label{fig:spinDyn} (a) Magnetic excitation dispersion in the (H~0~4.67$\pm$0.1) direction from neutron inelastic scattering measurements performed on CAMEA at T = 1.5 K. (b) Comparison with S(Q,$\omega$) calculated from Monte Carlo simulations, using the model given in table \ref{tab:param}. (c-d) Constant-Q energy scans measured on IN20 at different temperatures, revealing higher energy magnetic excitations. (e)  Energy scan measured at the Brillouin zone boundary on IN12 showing several magnetic modes fitted by Lorentz functions. The orange dots in panels (a-b) result from such fits. }
\end{figure}

We can tentatively ascribe the highest temperature peak in $\chi''$ to the reversal/creation of chain-like defects whose presence in the collinear pattern allows the system to accommodate the incommensurate magnetic structure, as shown in Figure \ref{fig:spinDyn}(a). The lowest temperature peak is characterized by a longer relaxation time and a smaller energy barrier, suggesting that it is related to domain dynamics. It is worth noting that beyond these two well-defined processes, $\chi''$ presents a broad non-zero background, which indicates the existence of a large number of relaxation channels, with a huge relaxation time distribution. 
These complex dynamics also manifest in magnetization curves as a function of magnetic field at very low temperature (see Figure \ref{fig:chi_ac}(c)). At 1.2 K, our measurements with ${\bf H} \parallel {\bf b}$ recover the magnetization plateau at about 1/3 of the saturated magnetization between 0.1 and 0.55~T, previously reported in higher temperature measurements \cite{Regnault1977, Zhong2020, Halloran2023}. 
When decreasing further the temperature at 500 mK and below, i.e. below the higher temperature $\chi''$ peak, magnetization steps occur at intermediate magnetic field values, whose number increases with temperature lowering. In addition, the step positions depend on the field sweeping direction and rate (See Supp. Mat. \cite{supmat}). This behavior is characteristic of small magnetic avalanches which often occur at low temperature and are usually characteristic of domain reorganizations. In that sense, these steps may be the signature under applied magnetic field of the processes observed in the ac susceptibility. Finally, the hysteresis opening, which was only observed up to the plateau at higher temperature, is present up to saturation at 500 mK and below (see inset of Fig. \ref{fig:chi_ac}(c)), indicating the existence of metastable states over the whole field range. 
Macroscopic magnetic measurements at very low temperature thus provide new experimental evidences for a magnetic ground state encompassing magnetic defects associated with thermally activated processes. In the proposed picture, this disorder is intrinsically related to the fact that the observed incommensurate propagation vector describing the magnetic order is not compatible with a long-range $(uudd)$ arrangement of collinear spins.



To gain more insight on the microscopic parameters leading to this original magnetic state, the spin excitations were measured in zero-field using neutron inelastic scattering \cite{supmat}. 
Figure \ref{fig:spinDyn}(a) displays the spin wave dispersion of BCAO measured on CAMEA along the (H~0~4.67) reciprocal space direction at T~=~1.5~K in its ordered phase. 
A well defined mode is observed, with a rather flat bottom above an energy gap of about 1.45 meV at the magnetic Brillouin zone center, of the order of the highest energy gap determined from ac susceptibility. This mode disperses to reach about 3 meV at the zone boundaries where it separates into two branches while getting broader in energy. At higher energy, between 3.5 and 6 meV, a broad signal can also be observed in the continuity of the lower spinwave branches, as highlighted in the constant-Q cut measured on IN12 (Fig. \ref{fig:spinDyn}(e) and orange dots in panel (a)). The absence of dispersion along the c-axis confirms the 2-dimensional nature of the magnetic order \cite{supmat}. Consistent with other neutron studies \cite{Halloran2023}, another weakly dispersing mode is visible around 15~meV, as shown in the constant-Q cut (Fig. \ref{fig:spinDyn}(c)) measured on IN20. It disappears above $T_N$ thus confirming its magnetic nature. A broad excitation is also detected around 30 meV which is not correlated with T$_N$, but disappears at 300~K (see panel (d)). This is attributed to the spin-orbit excitation corresponding to the $j_{\rm eff}$=3/2 multiplet, usually observed in this energy range in other cobaltates \cite{Songvilay2020,Yuan2020,Elliot2021,Yao2022}. It confirms the $j_{\rm eff}$=1/2 ground state picture in BCAO and the assignment of the low-energy spectrum to excitations within this manifold. One intriguing feature of the spin waves resides in the absence of a Goldstone mode stemming from the incommensurate magnetic Bragg positions. The global gap, confirmed by another neutron study along other directions \cite{Halloran2023}, is not compatible with a helical arrangement of spins with an XY anisotropy in the ab-plane, hinting for the presence of an additional anisotropy within this easy-plane. 


Let us note that the origin of this additional in-plane anisotropy which aligns the spins along a preferred direction close to the $b$-axis, cannot be due to single-ion physics. Indeed, in BCAO, the distorted CoO$_{6}$ octahedra are in a trigonal antiprismatic configuration, with a conserved local $C_{3}$ symmetry at the Co$^{2+}$ ions sites (Wyckoff position 6c) \cite{supmat}. Therefore, from symmetry arguments, the single-ion anisotropy in BCAO is purely easy-plane since no in-plane easy-axis anisotropy that would break the $C_{3}$ symmetry is allowed  \cite{Regnault2018,Halloran2023,Das2021,Winter2022,Liu2023,Lee2024}. 
To account for an additional in-plane anisotropy, other ingredients must then be considered, which naturally leads to the possibility of bond-dependent interactions.

To proceed with the modeling of the experimental neutron inelastic data, we considered the generic $JK\Gamma\Gamma'$ Kitaev-Heisenberg matrix involving bond-dependent interactions \cite{Liu2021,supmat}. In first approximation, we only considered bond-dependent anisotropic interactions for the first nearest-neighbors as it was previously suggested that the contribution of such interactions for further neighbors should be weak \cite{Elliot2021,Halloran2023, Maksimov2022, Liu2023, Samanta2024}. In the local $xyz$ frame of the octahedra (see Fig.\ref{fig:chi_ac}(a)), the symmetry allowed matrix terms have this form:

\[ H^{z}_{(1)} = 
\begin{pmatrix}
J-\eta & \Gamma' & \Gamma_1\\
\Gamma' & J+\eta & \Gamma_2\\ 
\Gamma_1  & \Gamma_2 & J+K
\end{pmatrix}
 \]
expressed for the $z$-bond, where, among the six independent parameters, $J$ is the isotropic Heisenberg coupling, $K$ is a bond-dependent (Kitaev) interaction and $\Gamma$ and $\Gamma'$ are two off-diagonal bond-dependent symmetric exchange couplings. For simplicity, we fix $\eta = 0$ and $\Gamma=\Gamma_{1} = \Gamma_{2}$, thus neglecting the tiny octahedral distortion resulting in the loss of the local 2-fold axis \cite{Liu2023}. Our model also includes Heisenberg interactions beyond first neighbors up to the fourth ones (Fig. \ref{fig:chi_ac}(a)) and an XXZ anisotropy ($\Delta = 0.5$), to be consistent with the single-ion physics of Co$^{2+}$ in BCAO \cite{supmat}.

\begin{figure}
 \includegraphics[scale=0.34]{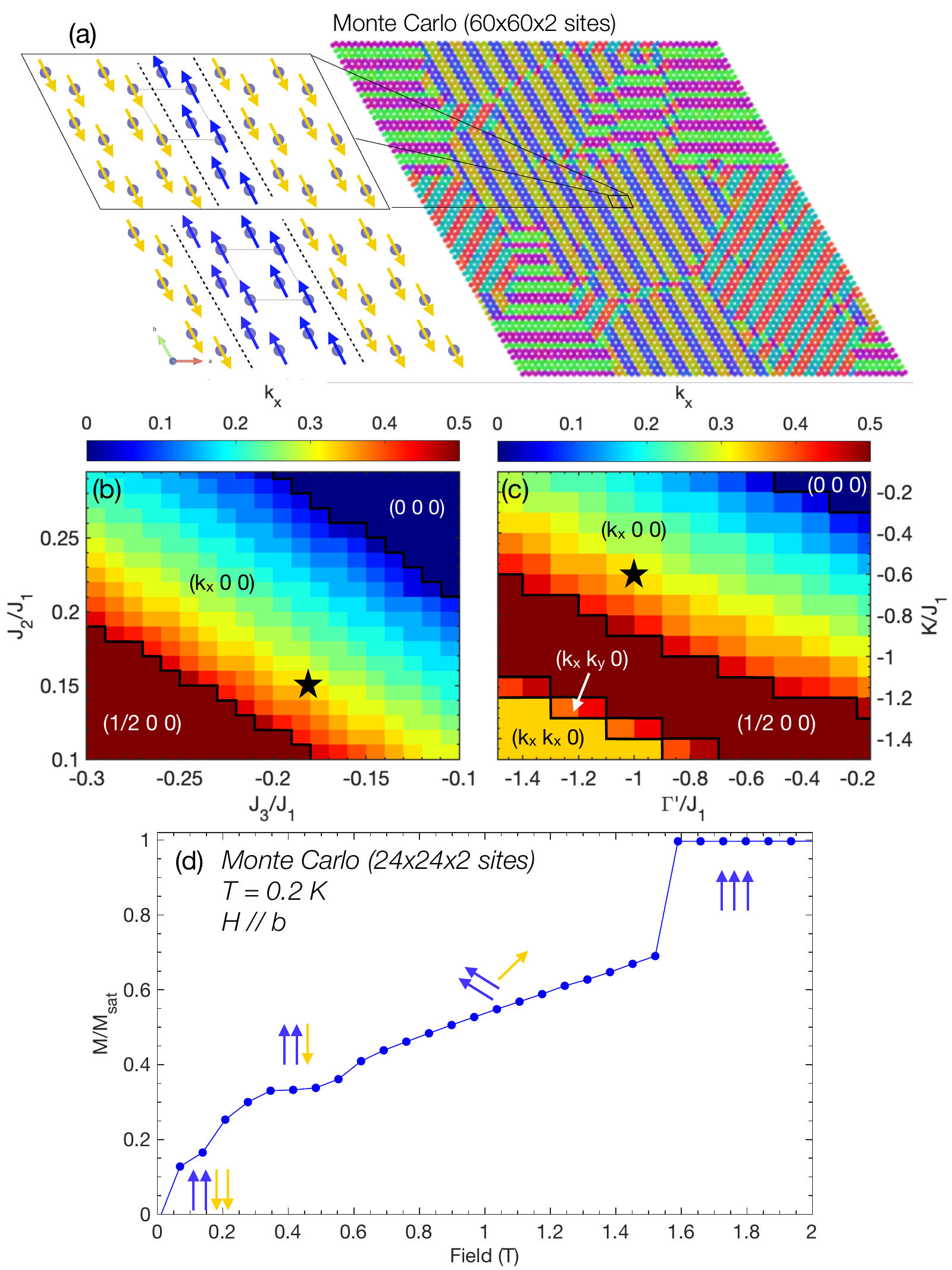}
 \caption{ \label{fig:MC} Monte-Carlo simulations using the model given in Table \ref{tab:param}: (a) spin structure where the colors represent different orientations of the spins, which are antiparallel within each three 120$^{\circ}$-domains. The inset illustrates the defective $uudd$ pattern for one domain and the flipping of a chain defect. (b-c) Magnetic phase diagrams calculated as a function of $J_{2}$ and $J_{3}$ (b) and $K$ versus $\Gamma'$ at fixed $\Gamma/J_{1}$ = -0.27 (c) using the Luttinger-Tisza method. The colors encode the ground state propagation vector component $k_x$. The interactions corresponding to the best model (see Table \ref{tab:param}) are indicated by the black stars. (d) Calculated magnetization curve as a function of magnetic field applied along b with a sketch of the spin arrangements of the different phases.}
\end{figure}

In order to find a spin wave model that reproduces the excitation dispersion, the parameter space was carefully explored by varying all the parameters. First, we determined regions in the parameter space that stabilise a ($k_{x}$~0~0) magnetic phase with $k_{x}$ close to the incommensurate value, by calculating phase diagrams using the Luttinger-Tisza method \cite{LTmethod}. Linear spin wave calculations were then performed using selected sets of parameters. Those giving the best agreement with the experimental spin wave dispersion were tested using Monte Carlo algorithm in order to check if they yield (i) a 1/3-magnetization plateau as a function of magnetic field ii) a $uudd$ spin arrangement disrupted by magnetic defects. Indeed linear spin wave calculations are performed on fully periodic ($k_{x}$=0.25) and ordered spin arrangements contrary to Monte Carlo method which is better adapted for ill-ordered magnetic states. Considering the above-mentioned constraints, several models are in close agreement with the experimental dispersion but only a few of them are also consistent with the disordered spin structure and 1/3-magnetization plateau within the observed magnetic field range. 

\begin{table}[t]
\caption{\label{tab:param} Set of interactions (in meV) corresponding to the best agreement with the magnetic structure, the magnetization $vs$ H, and the dispersion of magnetic excitations in BCAO. Plus/minus signs correspond to antiferromagnetic/ferromagnetic interactions.}
\begin{ruledtabular}
\begin{tabular}{lllllll}
$J_1$ & $J_2$ & $J_3$ & $J_4$ & $K$ & $\Gamma$ & $\Gamma'$ \\
 \hline
  -3.3(1) & -0.50(2) &  0.60(5) & 0.15(2) & 2.0(15) &  -0.9(1) & 3.3(1)\\
\end{tabular}
\end{ruledtabular}
\end{table}

Figures \ref{fig:spinDyn}(b) and \ref{fig:MC} show the results of the model yielding the best agreement with the experimental data (see \cite{supmat} for other models). The set of parameters given in table \ref{tab:param} reproduces well the low-energy excitation dispersion, the energy gap and the diffuse intensity above 4 meV at the zone boundaries (Fig. \ref{fig:spinDyn}(b)). It also generates optical branches although at an energy ($\approx$7 meV) lower than the observed higher magnon one around 15~meV (Fig. \ref{fig:spinDyn}(c)). Most importantly, as shown in Fig. \ref{fig:MC}(a), the stabilized spin structure corresponds to ferromagnetic chains with spins aligned along the $b$-axis, which are stacked along the a$^*$ direction in a $uudd$ sequence including defective chains, resulting in an average incommensurate propagation vector equal to 0.27 as observed experimentally. Finally the calculated magnetization curve qualitatively reproduces the 1/3-magnetization plateau above H$_{th1}$~=~0.2 T (Fig. \ref{fig:MC}(d)) observed in the experiment at H$_{c1}$~=~0.1 T (Fig. \ref{fig:chi_ac}(c)) with a $uud$ spin arrangement. Note that, in the calculation, between 0.6 and 1.6~T, the calculations favor a ``spin flop" configuration with the antiferromagnetic spin component perpendicular to the field, as also reported in another theoretical work \cite{Maksimov2023}. This additional phase could explain the non zero slope observed experimentally on the plateau and will deserve further investigations. 




The model we propose includes an antiferromagnetic Kitaev term $K$, which is not, however, the dominant interaction \cite{Liu2023,Lee2024}, thus contrasting with the recently studied delafossite-type honeycomb cobaltates \cite{Songvilay2020, Kruger2023}. 
It also implies significant off-diagonal $\Gamma$ and $\Gamma'$ terms, the latter being as large as the Kitaev term. Other Hamiltonians have been proposed in the literature to explain BCAO magnetic properties \cite{Halloran2023,Liu2023,Maksimov2022,Lee2024,Samanta2024}, displaying a rather large distribution of parameters. Some of these models are obviously insufficient since they predict a helical ground states \cite{Halloran2023,Lee2024}. Most of them agree with a large ferromagnetic $J_1$, a weaker antiferromagnetic Kitaev $K$ interaction and a non-negligible antiferromagnetic $J_3$ interaction, as in our results. We considered in addition second and fourth neighbors Heisenberg interactions. The latter, although weak, is essential in our semi-classical approach to stabilize the $uudd$ defective magnetic ground state, as well as shaping the Brillouin zone center excitation and stabilizing a 1/3 magnetization plateau. Note that a $uudd$ ground state has been predicted in other studies by adding either quantum fluctuations \cite{Jiang2023,Lee2024} or anisotropic interactions on the 3rd neighbor path \cite{Maksimov2022}. These additional parameters cannot be excluded. Quantum effects could account from some broadening of our measured spinwave excitations (which could also be attributed to disorder) and question the nature of the magnetic excitations beyond linear spin waves. Considering, on the other hand, anisotropic 3rd-neighbor interactions could improve the high-energy part of the magnetic excitations, in particular yielding the 15 meV mode \cite{Maksimov2023}. However, it is to be emphasized that none of the cited models were shown to reproduce the spin waves dispersion in zero-field (except from the spin gap), whereas our set of parameters accounts rather well simultaneously for the low-energy magnetic excitations, the 1/3 magnetization plateau and the peculiar ground state. Finally, note that the propagation vectors corresponding to our Hamiltonian found by the Luttinger-Tisza method is closer to 1/3 than to 0.27 obtained from Monte-Carlo calculations. Moreover, there is a locking of the propagation vector (1/3,0,0) measured by neutron diffraction on the 1/3 magnetization plateau after decreasing the magnetic field almost down to zero \cite{Regnault1979}. Both results therefore point to the proximity in energy of the uud phase from the one we have described above as the ground state. 



This intrinsically disordered magnetic structure emanates from a delicate balance between the competing interactions at play and the in-plane easy-axis anisotropy. Some hallmarks of this ground state have been observed in other systems, in particular with Co$^{2+}$, displaying strong Ising-like anisotropy and frustration, such as the anisotropic next-nearest-neighbor interactions (ANNNI) model yielding a devil's staircase physics or the triangular lattice. In some of these cobalt oxides, multiple magnetization steps as a function of field, frequency-dependent ac susceptibility and complex phase diagrams including short-range orders have been reported \cite{Selke1988,Kamiya2012,Matsuda2015,Abe2024,Hardy2004,Hardy2004b,Maignan2000,Jo2009,Edwards2020}. BCAO represents here an original example were the Ising-like spin anisotropy stems directly from anisotropic interactions. Note that similar signature of magnetic defects have also been recently reported in the sister material BaCo$_{2}$(PO$_{4}$)$_{2}$ via magnetostriction measurements at low temperature \cite{Wang2023} and in another cobalt honeycomb lattice compound, Na$_{3}$Co$_{2}$SbO$_{6}$. In the latter, the measured strong in-plane anisotropy may reflect significant contribution from bond-dependent interactions \cite{Li2022}. 

In conclusion, our study highlights, for the first time, the existence of an original ill-ordered magnetic state in BCAO, with further evidence for chain-like magnetic defects, and establishes the role of anisotropic bond-dependent interactions in stabilizing such exotic spin texture. Our Monte Carlo calculations have indeed shown that a spin Hamiltonian including anisotropic bond-dependent interactions was able to describe the magnetic structure, magnetization data and excitations in BCAO, thus providing a comprehensive understanding of this material.  Beyond Kitaev physics, this highlights the relevance of such contributions in the description of 3d transition metal magnetism and opens new prospects for exploring new exotic magnetic states in these materials.

\begin{acknowledgments}
We are grateful to L.P. Regnault and J. E. Lorenzo for lending us the samples and for fruitful discussions. We thank J. Debray for the sample orientation, the SERAS teams (Institut N\'eel) for manufacturing the sample mount, C. Paulsen for allowing us to use his SQUID dilution magnetometers and D. Jegouso for the optical spectroscopy measurements. We acknowledge funding from ANR (ANR-21-CE30-0034) and the support of the  Institut Laue Langevin and 2FDN in providing the access to research facilities used in this work. We acknowledge the Paul Scherrer Institute for the allocated beamtime on CAMEA, The neutron scattering data collected on IN12 are available at \cite{doiILL}.
\end{acknowledgments}

%

\end{document}